\documentclass[a4paper,UKenglish,cleveref, autoref, thm-restate]{lipics-v2021}

\pdfoutput=1 
\hideLIPIcs  

\usepackage{pgfplots}
\pgfplotsset{compat=1.18}
\usepackage{tikz}
\usepackage{subcaption}
\usepackage{tabularx}
\usepackage{threeparttable}

\newboolean{showcomments}
\setboolean{showcomments}{true} 
\ifthenelse{\boolean{showcomments}}
{\newcommand{\nb}[2]{
		\fcolorbox{gray}{yellow}{\bfseries\sffamily\scriptsize#1}
		{$\blacktriangleright$#2$\blacktriangleleft$}
	}
	
}
{\newcommand{\nb}[2]{}
	 
}

\title{Why Do LLMs Fail at OCL Generation? A Graph Reasoning Perspective}

\author {Hamza Attarwala} {Polytechnique Montréal, Montréal, QC, Canada \and \url{https://hamzagitx786.github.io/}} {hamza-salim.attarwala@etud.polymtl.ca} {https://orcid.org/0009-0001-5442-5210} {}

\author{ Moataz Chouchen}{ Concordia University, Montréal, Canada \and \url{https://www.concordia.ca/faculty/moataz-chouchen.html} }{ moataz.chouchen@concordia.ca }{ https://orcid.org/0000-0002-1134-1324 }{}

\author {Mohammad Hamdaqa} {Polytechnique Montréal, Montréal, QC, Canada \and \url{https://www.polymtl.ca/expertises/en/hamdaqa-mohammad}} {mhamdaqa@polymtl.ca} {https://orcid.org/0000-0003-4927-2755} {}

\author {Omar Alam} {Trent University, Peterborough, ON, Canada \and \url{https://omaralam.org/}} {omaralam@trentu.ca} {https://orcid.org/0000-0003-3973-9147} {}

\authorrunning{Attarwala et al.}



\nolinenumbers 

\EventEditors{John Q. Open and Joan R. Access}
\EventNoEds{2}
\EventLongTitle{42nd Conference on Very Important Topics (CVIT 2016)}
\EventShortTitle{CVIT 2016}
\EventAcronym{CVIT}
\EventYear{2016}
\EventDate{December 24--27, 2016}
\EventLocation{Little Whinging, United Kingdom}
\EventLogo{}
\SeriesVolume{42}
\ArticleNo{23}

\begin{document}

\maketitle

\begin{abstract}
\textbf{Background.} Large Language Models (LLMs) are increasingly used to generate Object Constraint Language (OCL) constraints from natural language specifications and UML class diagrams. However, existing work mainly focuses on improving accuracy, with limited understanding of why these models fail. \textbf{Aims.} This study investigates the underlying causes of LLM failures in OCL generation, framing the task as a graph reasoning problem over UML class diagrams. \textbf{Method.} We conduct an empirical evaluation using the PathOCL dataset across six state-of-the-art LLMs. We analyze the impact of UML structural properties (e.g., navigation depth and model complexity), lexical similarity, prompt ordering strategies, and graph-aware prompting on OCL correctness. \textbf{Results.}We find that OCL generation performance significantly degrades with increasing navigation depth and structural complexity. Lexical similarity has limited influence, while textual ordering of UML elements affects performance. Graph-based prompting yields partial improvements but does not eliminate structural reasoning errors. \textbf{Conclusions.}  OCL generation is primarily constrained by graph reasoning limitations rather than purely linguistic factors. These results highlight structural reasoning as a key bottleneck for current LLMs in model-driven engineering tasks.
\end{abstract}

\vspace{-0.2cm}
\section{Introduction}

Large Language Models (LLMs) are increasingly used to automate software engineering tasks that translate natural language (NL) specifications into formal representations. One important task is the generation of Object Constraint Language (OCL) expressions from NL specifications and UML class diagrams \cite{OMG_OCL_2014, abukhalaf2023codex, bajwa2010ocl, cabot2022combining}. Accurate OCL generation could reduce manual effort in model-driven engineering and improve specification consistency \cite{cabot2012object, cabot2014verification}. However, recent studies show that LLMs frequently generate erroneous OCL constraints even when advanced prompting and fine-tuning techniques are used \cite{abukhalaf2023codex, pan2024generative, li2025optimizing}.

Prior work has mainly focused on improving OCL generation through prompt engineering, fine-tuning, and dataset construction \cite{abukhalaf2023codex, abukhalaf2024pathocl, pan2024generative}, with most efforts targeting incremental performance improvements. As a result, they provide limited insight into \emph{why} LLMs fail. In particular, little attention has been given to the role of structural reasoning over UML models, despite its fundamental importance for correct OCL generation \cite{tan2010ocl}.

A UML class diagram can naturally be represented as a graph, where classes correspond to nodes and associations correspond to edges \cite{holscher2006translating, ziemann2005uml}. Generating OCL constraints often requires multi-hop navigation across this graph to identify relevant entities and relationships. This suggests that OCL generation is not merely a language translation task, but also a graph reasoning problem.

Recent graph reasoning benchmarks show that LLMs struggle with multi-step reasoning over graph structures, especially as graph depth and complexity increase \cite{wang2023can}. These findings motivate the following question: \emph{Do similar structural reasoning limitations explain failures in LLM-based OCL generation using UML models?}

To investigate this question, we study how UML structural properties and non-structural lexical cues influence LLM-generated OCL constraints. Specifically, we address the following research questions:

\vspace{0.3em}

\noindent\textbf{RQ1. How do UML structural properties influence LLM-based OCL generation?}
\begin{itemize}
\item \textbf{RQ1.1.} How does structural navigation depth affect OCL correctness?
\item \textbf{RQ1.2.} Does OCL generation degrade with increasing UML structural complexity?
\end{itemize}

\noindent\textbf{RQ2. To what extent do lexical cues influence LLM-based OCL generation?}
\begin{itemize}
\item \textbf{RQ2.1.} How does lexical similarity between NL specifications and UML elements influence generated OCL?
\item \textbf{RQ2.2.} Does the ordering of classes and associations in the UML model as a textual representation affect OCL generation?
\end{itemize}

\noindent\textbf{RQ3. How can OCL generation be improved, and which errors persist?}
\begin{itemize}
\item \textbf{RQ3.1.} Do advanced prompting strategies (e.g., Chain-of-Thought, Few-shot, Build-a-Graph) improve OCL generation quality?
\item \textbf{RQ3.2.} What types of OCL generation errors occur across different conditions during this empirical study?
\end{itemize}

Based on these questions, this paper makes the following contributions:

\begin{itemize}
\item We conceptualize UML-based OCL generation as a graph reasoning problem and relate OCL failures to known LLM limitations in graph reasoning.
\item We conduct an empirical study to evaluate the effects of structural complexity, navigation depth, lexical similarity, and prompt design on OCL generation.
\item We evaluate advanced prompting strategies, including Chain-of-Thought and Build-a-Graph prompting, for improving OCL generation.
\item We identify and analyze recurring OCL generation errors, including invalid navigation, incorrect use of OCL language, hallucinated content, and syntactically valid but semantically invalid OCL expressions.
\end{itemize}

Our results show that OCL generation performance degrades with increasing UML structural depth and complexity, consistent with prior graph reasoning studies. We find limited evidence for a strong lexical bias explanation: LLMs do not appear to rely solely on superficial lexical similarity, although misleading class-level lexical cues can still negatively affect generation in some cases. Advanced prompting strategies provide only partial improvements. Overall, this study shifts the focus from improving OCL generation accuracy toward understanding the underlying causes of LLM failures in model-driven engineering tasks.
\vspace{-0.2cm}
\section{Motivation}

Recent work has shown that Large Language Models (LLMs) struggle with reasoning tasks involving graph-structured data, when the graph is expressed entirely in natural language \cite{wang2023can, ge2025can, li2025graphotter, peng2025rewarding}. Across graph reasoning benchmarks, LLMs often generate fluent intermediate explanations while failing to correctly reason about structural relationships such as connectivity, traversal, and path optimization. These findings are particularly relevant to UML-based OCL generation, since generating a valid OCL constraint frequently requires navigating multiple associations and maintaining consistency with the underlying UML graph structure.

Listing \ref{lst:nlgraph-exp1} shows one representative example is reported in the NLGraph benchmark \cite{wang2023can}, which evaluates LLMs on graph reasoning tasks formulated in natural language. In this task, an LLM is asked to identify whether a path between two nodes exist.

\begin{lstlisting}[caption={Path reasoning failure from \cite{wang2023can}}, label={lst:nlgraph-exp1}, mathescape=true]
Question: "Is there a path, in this undirected graph, that exist 
between node 4 and node 5? If yes, give the path. "
Graph connections:
$0 \longrightarrow 4$, $1 \longrightarrow 6$, $3 \longrightarrow 5$, $2 \longrightarrow 6$, $1 \longrightarrow 3$, $2 \longrightarrow 5$, $5 \longrightarrow 6$, $1 \longrightarrow 2$
Correct path:
No path exist that connect Node 4 and Node 5 
LLM output:
Yes, a path between Node 4 and Node 5 does exisit:
$4 \longrightarrow 0 \longrightarrow 6 \longrightarrow 2 \longrightarrow 5$
\end{lstlisting}

The path generated by the model is not valid. The example illustrates an important limitation: the model fails to correctly captures local graph connectivity, and fails to reason about the graph structure. Other work additionally demonstrates that LLM performance is sensitive to how graph information is presented in text. For example, Ge et al. \cite{ge2025can} show that simply reordering graph edges in the prompt can substantially change reasoning accuracy, even when the underlying graph remains identical.

\begin{figure}
    \centering
    \includegraphics[width=0.6\linewidth]{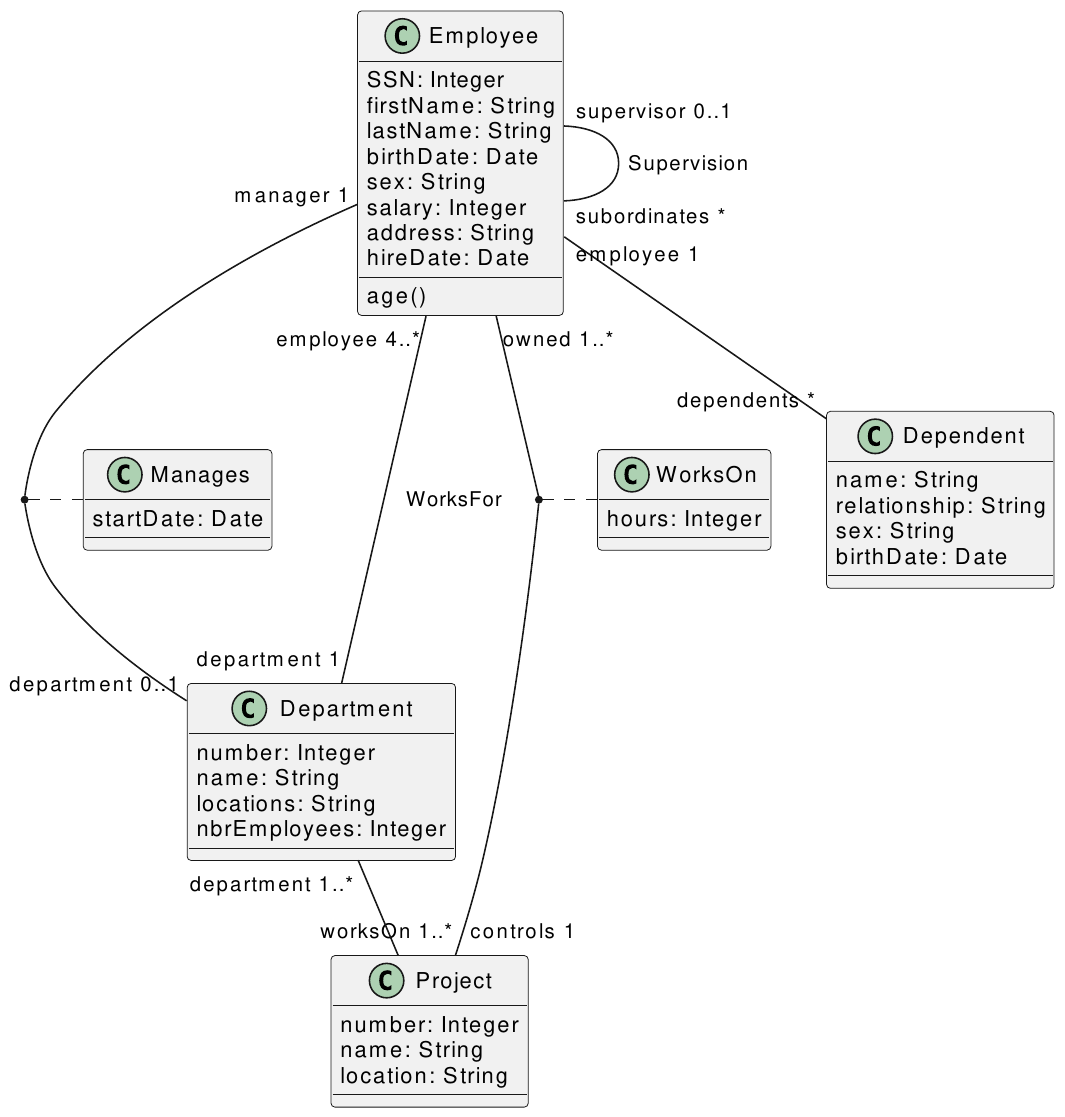}
    \caption{A class diagram from the EmpoymentAgency Domain}
    \label{fig:class_diagram_emp}
\end{figure}

These limitations are directly reflected in OCL generation from UML class diagrams. Although OCL is a formal specification language, producing correct constraints requires reliable navigation over the underlying UML graph, including identifying valid association paths, preserving navigability constraints, and maintaining consistency across multi-step traversals between classes. Listing~\ref{lst:incorrect_OCL_expression} illustrates this issue through an example generated by \emph{Claude Sonnet 4.5} for the natural language specification: \textit{Only department managers can work less than 5 hours on a project}, based on the UML class diagram in Figure~\ref{fig:class_diagram_emp}, where the resulting OCL expression is incorrect due to failures in maintaining valid structural navigation. 

\begin{lstlisting}[caption={OCL expression generated by LLM}, label={lst:incorrect_OCL_expression}]
context WorksOn inv: self.hours < 5 implies 
self.employee.department.manager = self.employee
\end{lstlisting}

The generated constraint incorrectly attempts to navigate through an association named \verb|employee| from the \verb|WorksOn| context class. Similar to the NLGraph example, the model produces syntactically plausible output while failing to preserve the underlying graph structure of the problem domain.

Motivated by these findings, this work investigates whether structural properties of UML class diagrams systematically influence OCL generation performance. Prior graph reasoning studies show that LLM performance degrades as reasoning paths become longer and graph structures become increasingly large or densely connected \cite{wang2023can, rameshkumar2025reasoning}. Accordingly, \textbf{H1} investigates whether increasing UML navigation depth reduces OCL generation accuracy, while \textbf{H2} examines whether increasing UML structural complexity similarly degrades performance. Existing work also demonstrates that graph reasoning performance is sensitive to the textual ordering of graph elements \cite{ge2025can}, motivating \textbf{H3}, which investigates whether different textual order of UML classes and associations with the prompt to the LLM influences the generated OCL constraints. 

Furthermore, prior studies suggest that LLMs sometimes rely on superficial lexical cues, such as mention frequency, rather than faithfully reasoning over graph structure \cite{han2025reasoning, wang2023can}. This motivates \textbf{H4}, which examines whether prominently mentioned entities in natural language specifications disproportionately influence generated OCL constraints, even when structurally irrelevant. Finally, previous work indicates that graph-oriented prompting strategies, including stepwise reasoning and intermediate graph construction, can partially mitigate structural reasoning failures \cite{wang2023can}. Consequently, \textbf{H5} investigates whether graph-aware prompting strategies improve OCL generation performance.

\vspace{-0.2cm}
\section{Methodology}

\noindent\textbf{Data.} The empirical evaluation conducted in this study uses the dataset introduced in PathOCL \cite{abukhalaf2024pathocl}. The original dataset consists of 15 UML class models originating from multiple real-world application domains, including airport systems, healthcare, and employment management. For each UML model, the dataset provides three main artifacts: (1) the UML class model itself, represented in PlantUML format, (2) corresponding natural language specifications describing system constraints, and (3) reference gold OCL constraints associated with those specifications.

Since the gold OCL constraints were required for evaluation, two UML models were excluded from the study because they lacked corresponding gold OCL annotations. After filtering, the final dataset comprised 13 UML class models and 115 natural language specifications paired with gold OCL expressions, substantial enough for an empirical study \cite{polo2024tinybenchmarks, pacchiardi2024100, huang2025minilongbench}. All UML models were represented in textual form using the PlantUML format \cite{plantuml}. The gold OCL constraints were used as ground truth references during evaluation. In particular, they were used to validate the correctness of the OCL constraints generated by the evaluated LLMs through test-instance-based verification procedures.

\noindent\textbf{Evaluated Language Models.} This study evaluated six large language models (LLMs), covering both closed and open-source models. The evaluated models included Claude Sonnet 4.5 \cite{claude2025}, GPT-5 \cite{gpt5}, Llama 4 Scout \cite{meta2024llama4}, Gemini 2.5 Pro \cite{gemini2025}, DeepSeek Chat V3.1 \cite{deepseek2024}, and Grok 4 Fast \cite{grok2025}. For all experiments, the temperature was fixed at 0.1 and the maximum generation length was set to 1500 tokens, allowing sufficient output capacity for OCL generation. Additionally, to account for the stochastic nature of LLM decoding, each input specification was evaluated over 10 independent runs per model under identical prompting and decoding settings. This repeated-sampling protocol was used to reduce variance in generation quality and obtain a more stable estimate of model behaviour.

The 10 generated OCL outputs per specification were aggregated into a single prediction using a deterministic majority-based selection strategy. First, outputs were normalized by removing formatting and whitespace variations, and syntactically equivalent OCL expressions were grouped together. The final prediction was selected as the most frequently occurring OCL expression across the 10 runs. In cases where no unique majority emerged, ties between candidate expressions were resolved deterministically by selecting the first valid OCL expression in the ordered sequence of runs. Outputs that failed to produce syntactically valid OCL expressions were excluded prior to aggregation.

\begin{figure}
    \centering
    \includegraphics[width=0.75\linewidth]{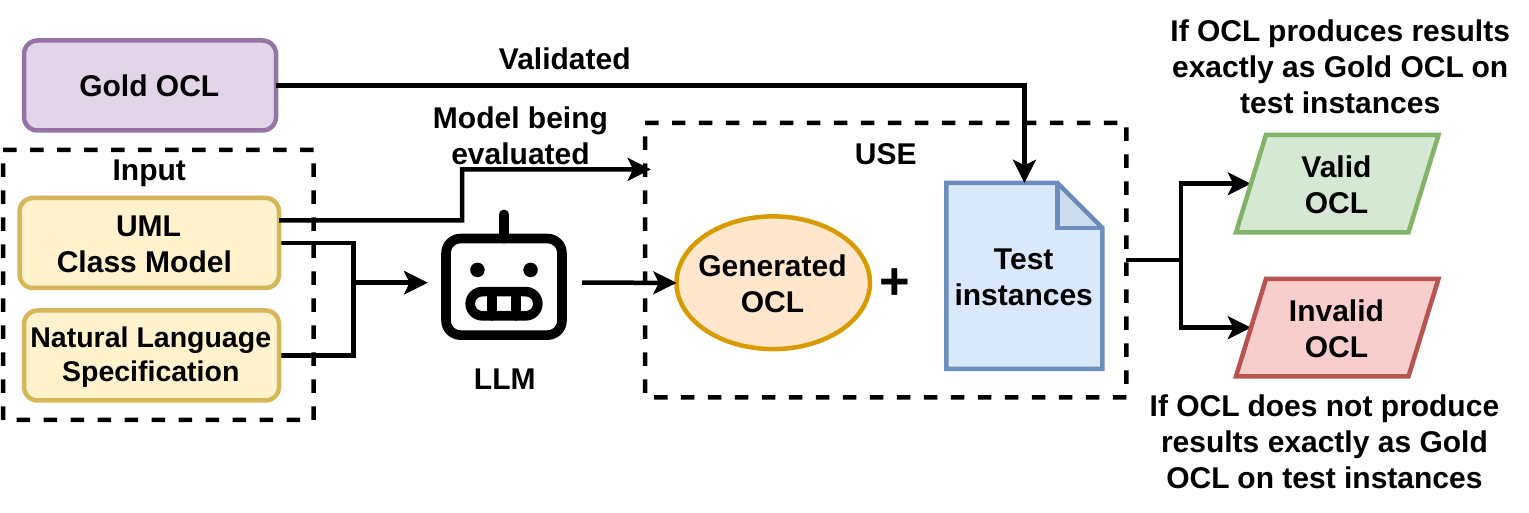}
    \caption{Process to categorize LLM-generated OCL expressions as valid or invalid}
    \label{fig:ocl_generation}
\end{figure}

\noindent\textbf{OCL Generation and Evaluation Procedure.}\label{OCL_correctness_evaluation} The OCL generation and evaluation procedure used in this study is illustrated in Figure \ref{fig:ocl_generation}. For each UML model in the dataset, the model in the text format and its associated natural language specification are provided as inputs to the LLM. Based on these inputs, the LLM generates an OCL expression intended to apply the constraints stated within NL specifications with respect to the provided UML model. The generated OCL expressions are then evaluated using the USE (UML-based Specification Environment) \cite{gogolla2007use} tool.

The evaluation process consists of two stages: syntactic validation and behavioural validation. In the first stage, the generated OCL expressions are checked for syntactic correctness using USE. This step determines whether the generated constraints conform to the grammar and structural rules of OCL. Any generated constraint that cannot be successfully parsed or executed within USE is classified as syntactically invalid.

In the second stage, the generated OCL expressions are evaluated behaviourally to determine whether they correctly capture the intended semantics of the NL specification. For each specification, two manually crafted \emph{object} test instances conforming to the corresponding UML model are created: (1) a positive test instance satisfying the specification, and (2) a counterexample violating the specification. To improve robustness, the test instances include not only straightforward satisfying and violating cases, but also structurally challenging scenarios designed to expose common OCL generation errors such as incorrect navigation paths, quantifier misuse, cardinality violations, and logical inconsistencies. Prior to evaluation, all test instances are validated against the gold OCL constraints to ensure they correctly represent the intended specification semantics.

Once validated, the same test instances are executed against the LLM-generated OCL expressions using USE. A generated OCL expression is considered behaviourally valid if it produces the same boolean evaluation outcomes as the corresponding gold OCL across all test instances; otherwise, it is classified as behaviourally invalid. This evaluation procedure is applied consistently across invariants, preconditions, and postconditions, enabling uniform comparison between generated and reference OCL constraints.

\noindent\textbf{Experimental Setup and Graph Construction.} To investigate how structural properties of UML class diagrams influence OCL generation, each UML model was first transformed into a graph-based representation. Since UML class diagrams naturally encode structural relationships between entities, representing the models as graphs enabled systematic analysis of graph-related properties relevant to OCL generation.

The UML models, represented in PlantUML format, were parsed using regular-expression-based extraction procedures. During this process, classes, attributes, operations, and associations were identified and converted into graph components. Specifically, UML classes were represented as graph nodes, while associations between classes were represented as edges. Attributes and operations associated with a class were stored as node-level properties. Additional metadata describing association types, multiplicities, navigability, and association classes were also preserved within the graph representation as edge properties.

This graph transformation enabled the extraction of structural properties required for the empirical analysis. In particular, breadth-first search (BFS) traversal was used to analyze connectivity and navigation relationships between classes. Using the generated graph representation, several structural metrics were computed for each UML model, including: 1) number of classes, 2) number of associations, 3) number of attributes, 4) number of operations, and 5) navigation depth between related classes. These graph-derived metrics were subsequently used as explanatory variables in the statistical analysis to investigate their relationship with OCL generation correctness.

\noindent\textbf{Statistical Analysis.} To analyze the relationship between UML structural properties and OCL generation correctness, this study employed Generalized Estimating Equations (GEE) \cite{hardin2002generalized} model. The dependent variable in the analysis was whether a generated OCL constraint was behaviourally correct according to the evaluation procedure described in Section~\ref{OCL_correctness_evaluation}.

A standard logistic regression model was considered, but not used because one of its key assumptions, namely the independence of observations, was violated in the experimental setting. Specifically, multiple natural language specifications were associated with the same UML class model. Since structural features such as the number of classes, associations, attributes, and graph connectivity were extracted at the UML-model level, multiple observations originating from the same UML model shared identical structural characteristics. Consequently, the generated samples were not statistically independent.

Hence, to account for this intra-model correlation, the GEE model was employed. GEE extends generalized linear models by explicitly modelling correlations among grouped observations while still estimating population-level effects. Using GEE allowed us to more reliably estimate the influence of structural UML properties on OCL generation performance while accounting for dependencies between specifications derived from the same UML model.
\vspace{-0.2cm}
\section{Results}

\subsection{RQ1: Effect of UML Structural Properties on OCL Generation}

\noindent\textbf{RQ1.1: Effect of Navigation Depth.}

To investigate whether deeper UML navigation affects OCL generation, we operationalize navigation depth from the reference OCL constraints. In OCL, a navigation expression corresponds to a sequence of association traversals starting from the context class of the UML model, where each role access in a \verb|self|-based expression represents one step in the UML graph. Navigation depth is defined as the maximum length of any such navigation chain in the gold OCL, capturing the extent of structural traversal required to express a constraint.
 
This metric captures the structural traversal required to express a constraint: shallow constraints involve only local attributes or directly associated classes, whereas deeper constraints require navigation across multiple related classes. Although we extract the metric from the OCL expression itself, it directly reflects the underlying UML graph structure, since each role access corresponds to traversal along an association in the class diagram.

To evaluate the effect of navigation depth on OCL generation correctness, the navigation-depth feature was incorporated as a predictor variable within the GEE model. Table~\ref{tab:nav-depth-descriptive} presents the observed correctness rates across different navigation depths. The results indicate a consistent decline in correctness as navigation depth increases. Constraints with navigation depths of 0 and 1 achieved correctness rates of 76.9\% and 71.9\%, respectively. In contrast, correctness decreased to 46.7\% for depth-2 constraints and further declined to 6.3\% for depth-3 constraints. These results indicate that OCL constraints requiring longer structural traversals are substantially more difficult for LLMs to generate correctly.

\begin{table}[t]
\centering
\caption{Effect of navigation depth on OCL correctness.}
\label{tab:nav-depth-results}

\begin{subtable}[t]{0.45\textwidth}
\centering
\footnotesize
\setlength{\tabcolsep}{6pt}
\caption{Correctness by navigation depth.}
\label{tab:nav-depth-descriptive}
\begin{tabular}{p{1.2cm}p{1cm}p{1.2cm}p{0.7cm}p{1cm}}
\hline
Navigation depth & Clusters & Generated & Correct & Correctness \\
\hline
0 & 18 & 360  & 277 & 76.9\% \\
1 & 54 & 1080 & 777 & 71.9\% \\
2 & 27 & 486  & 227 & 46.7\% \\
3 & 7  & 126  & 8   & 6.3\% \\
\hline
\end{tabular}
\end{subtable}
\hfill
\begin{subtable}[t]{0.45\textwidth}
\centering
\footnotesize
\caption{GEE regression results.}
\label{tab:nav-depth-gee}
\begin{tabular}{p{1.2cm}rrrr}
\hline
Predictor & Coef. & OR & SE & p-value \\
\hline
Navigation depth & -0.937 & 0.392 & 0.218 & 0.0031 \\
\hline
\end{tabular}
\end{subtable}

\end{table}

The GEE analysis further confirms that this relationship is statistically significant. Navigation depth produced a negative coefficient of $-0.937$, with an odds ratio of $0.392$ and a p-value of $0.0031$ (Table~\ref{tab:nav-depth-gee}). Since correctness was used as the binary outcome variable, the negative coefficient indicates that increasing navigation depth significantly decreases the probability of generating a behaviourally correct OCL constraint. Interpreted in terms of odds ratios, a one-standard-deviation increase in navigation depth corresponds to an approximate 60.8\% reduction in the odds of generating a correct OCL expression.

The observed effect remained stable across alternative correlation assumptions. When the GEE model was refitted using an independence correlation structure, navigation depth remained negative and statistically significant, with coefficient $-0.980$, odds ratio $0.375$, and $p=0.0021$. A similar pattern was also observed consistently across all evaluated LLMs, where navigation depth produced negative and statistically significant coefficients for each model individually.

The results provide strong evidence that navigation depth negatively affects LLM-based OCL generation performance. As the required structural traversal through the UML model increases, the likelihood of generating a correct OCL constraint decreases substantially. These findings are consistent with the graph-reasoning perspective motivating this study and suggest that many OCL generation failures are associated with difficulties in maintaining longer structural navigation paths over UML class graphs.

\noindent\textbf{RQ1.2: Effect of Structural Complexity.}

We define UML structural complexity as the size and connectivity of the UML class model that the LLM must reason over when generating an OCL constraint. In graph terms, structurally simple model contains fewer possible classes and navigation alternatives, whereas a structurally complex model contains more classes, more associations, and more model elements that may need to be selected, ignored, or navigated correctly. In this study, we operationalized structural complexity using three model-level metrics: the number of classes, the number of associations, and the total number of attributes and operations.

For each UML model, these metrics were extracted from the graph-based class diagram. Since all specifications belonging to the same UML model share the same structural-complexity values, we evaluated the relationship between each complexity metric and OCL correctness using a logistic GEE model with correctness as the binary outcome. To avoid instability caused by strong correlations among structural metrics, we fitted one GEE model per complexity metric. Each feature was standardized before fitting, so the coefficients represent the effect of a one-standard-deviation increase in the corresponding complexity measure.

Table~\ref{tab:structural-complexity-results} reports the results. The number of classes has a statistically significant negative effect on OCL correctness ($\beta=-0.514$, $p<0.0012$). The corresponding odds ratio is 0.598, indicating that a one-standard-deviation increase in the number of classes is associated with an approximately 40.2\% decrease in the odds of generating a correct OCL expression. Similarly, the number of associations also has a statistically significant negative effect ($\beta=-0.523$, $p<0.0024$), with an odds ratio of 0.593. This corresponds to an approximately 40.7\% decrease in the odds of correctness for a one-standard-deviation increase in associations.

\begin{table}[t]
\centering
\footnotesize
\caption{Effect of UML structural complexity on OCL correctness.}
\label{tab:structural-complexity-results}
\begin{tabular}{lrrrr}
\hline
Structural metric & Coefficient & Odds ratio & 95\% CI & p-value \\
\hline
Number of classes & -0.514 & 0.598 & [-0.756, -0.272] & $<0.0012$ \\
Number of associations & -0.523 & 0.593 & [-0.771, -0.274] & $<0.0024$ \\
Attributes and operations & -0.347 & 0.707 & [-0.779, 0.085] & 0.115 \\
\hline
\end{tabular}
\end{table}

In contrast, the number of attributes and operations does not show a statistically significant independent effect ($\beta=-0.347$, $p=0.115$). Although the coefficient is negative, the confidence interval crosses zero. This suggests that the amount of local class-level information alone is not sufficient to explain variation in OCL correctness. One possible reason is that attributes and operations are highly correlated with other model-size measures, especially the number of classes and associations, making their independent contribution harder to isolate.
\begin{figure}
    \centering
    \includegraphics[width=0.75\linewidth]{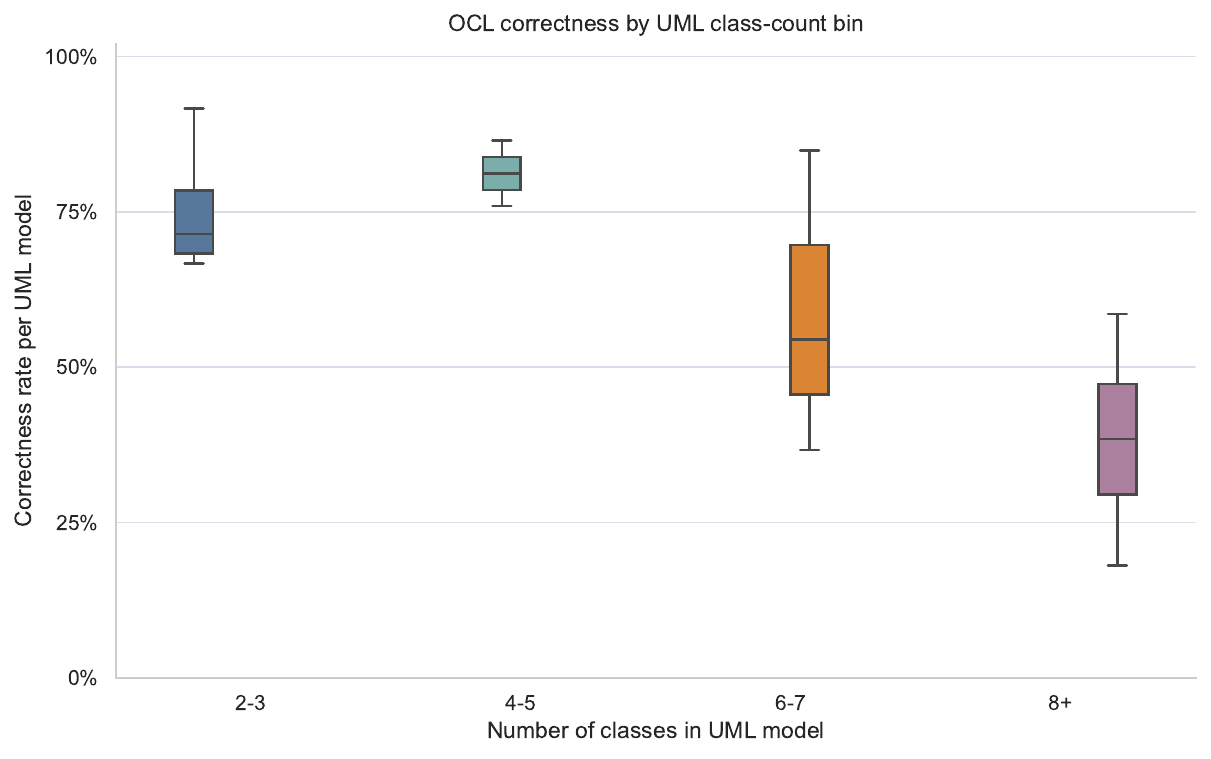}
    \caption{Distribution of OCL correctness across UML class-count bins.}
    \label{fig:box_plot}
\end{figure}
Figure~\ref{fig:box_plot} further illustrates the relationship between UML structural complexity and OCL generation correctness using class-count bins as a proxy for model size. The observed distributions are consistent with the GEE results, showing a clear decline in correctness as the number of classes increases. UML models containing two to three classes achieve an average correctness of approximately 75.3\%, whereas models containing eight or more classes achieve an average correctness of only 38.4\%, representing nearly a 49\% relative reduction.

The box plots further show that smaller UML models (0--4 classes) generally achieve higher median correctness scores, although with greater variability across specifications and prompting configurations. As model size increases, the median correctness decreases consistently across successive bins, while the distributions become increasingly concentrated near lower correctness values. In particular, the largest complexity bins (8+ classes) contain comparatively fewer high-performing observations, indicating that structurally larger UML models produce consistently poorer OCL generation performance. This supports the regression results and indicates that LLM performance decreases as the UML model becomes structurally larger and more connected.

\vspace{-0.4cm}
\subsection{RQ2: Influence of Lexical and Ordering Biases}

\noindent\textbf{RQ2.1: Lexical Similarity Effects}

This research question investigates whether lexical similarity between the natural language specification and UML model elements is associated with the performance of LLM-based OCL generation. To evaluate this hypothesis, we compare the natural language specification against UML elements that appear in the gold OCL constraint and against UML elements that do not appear in the gold OCL constraint. First, we extract the UML elements referenced by the gold OCL constraint, including the context class, referenced classes, terminal attributes, and navigated association roles. These elements form the \emph{gold} element set, representing the UML elements required to express the constraint. All remaining classes, attributes, and roles in the UML model are treated as \emph{non-gold} elements, representing potential distractor elements.

We then compute several lexical similarity features using a token-level Jaccard similarity metric. The gold-alignment features measure how strongly the natural language specification matches the UML elements referenced in the gold OCL constraint. Conversely, the non-gold similarity features measure similarity between the specification and UML elements not used in the gold constraint. We also compute a lexical trap score, defined as the maximum similarity between the specification and any non-gold UML element. Finally, we compute a distance-weighted trap score, which assigns higher weight to lexically similar non-gold classes that are structurally farther from the gold navigation path.

We hypothesize that if LLM-based OCL generation is influenced by lexical similarity, then higher similarity to gold elements is positively associated with correctness, whereas higher similarity to non-gold elements is negatively associated with correctness. Accordingly, gold-alignment features are expected to have positive coefficients, while lexical trap features are expected to have negative coefficients.

Table~\ref{tab:lexical-similarity-results} summarizes the GEE results for the main lexical-similarity hypothesis features. The overall lexical trap score was not statistically significant ($\beta=0.610$, $p=0.1134$), and its coefficient was positive rather than negative. Gold alignment was also not significant ($\beta=0.061$, $p=0.7681$). Similarly, the distance-weighted trap score was statistically significant but in the opposite direction from the hypothesis ($\beta=0.340$, $p=0.0458$), indicating that it does not support the proposed lexical-trap explanation. Similarity to non-gold attributes showed a negative trend, but it was not statistically significant ($\beta=-0.514$, $p=0.0927$). These results do not support the broad lexical-bias hypothesis. The expected pattern, where similarity to irrelevant UML elements consistently reduces correctness and similarity to gold elements improves correctness, was not observed. Across the main hypothesis tests, none of the four lexical-similarity features showed the expected statistically significant effect.

\begin{table}[t]
\centering
\footnotesize
\caption{GEE results for lexical-similarity hypothesis features.}
\label{tab:lexical-similarity-results}
\begin{tabular}{lrrrr}
\hline
Feature & Expected direction & Coefficient & Odds ratio & p-value \\
\hline
Lexical trap score & Negative & 0.610 & 1.841 & 0.1134 \\
Gold alignment & Positive & 0.061 & 1.063 & 0.7681 \\
Distance-weighted trap & Negative & 0.340 & 1.405 & 0.0458 \\
Similarity to non-gold attributes & Negative & -0.514 & 0.598 & 0.0927 \\
\hline
\end{tabular}
\end{table}

However, the broader model reveals a more nuanced result. Similarity to non-gold classes was statistically significant and negative ($\beta=-1.141$, odds ratio $=0.319$, $p=0.0006$). This suggests that class-level lexical distractors may still matter. When the natural language specification is lexically similar to classes that are not part of the gold OCL expression, the likelihood of generating a correct constraint decreases. This effect was also stable in the sensitivity analysis using an independence correlation structure ($\beta=-1.132$, odds ratio $=0.323$, $p=0.0007$).

Thus, the results provide limited rather than general support for lexical bias. We do not find evidence that LLMs rely primarily on lexical similarity across all UML element types. In particular, gold alignment, lexical trap score, and distance-weighted trap score do not behave as predicted. At the same time, the significant effect of non-gold class similarity indicates that lexical overlap with irrelevant class names can still interfere with OCL generation. This suggests that LLM failures cannot be explained by lexical similarity alone, and instead, lexical cues may interact with structural reasoning, especially when the model must choose among competing candidate classes in the UML graph.

Therefore, for RQ2.1, we conclude that lexical similarity is not a sufficient explanation for OCL generation errors. The results reject the broad lexical-trap hypothesis, but they leave room for a narrower class-level lexical distraction effect. This finding is consistent with the results from several graph reasoning paper \cite{wang2023can, zhang2024can} where LLMs appear to perform some reasoning over the UML model structure, but their generation can still be affected by misleading textual cues in specific cases.

\noindent\textbf{RQ2.2: UML Ordering Effects}

To investigate whether the textual ordering of UML elements influences OCL generation, we evaluated five strategies for linearizing UML class diagrams before providing them to the LLM. These include \emph{classes-first}, \emph{associations-first}, \emph{breadth-first search (BFS)}, \emph{depth-first search (DFS)} inspired by graph-reasoning studies \cite{ge2025can} and a proposed \emph{leaf-first} traversal. The classes-first and associations-first strategies grouped UML elements by type, whereas BFS, DFS, and leaf-first generated traversal-based ordering over the UML graph structure. The proposed leaf-first strategy begins from terminal classes and progressively moves toward more central classes, motivated by the observation that many OCL constraints ultimately reference terminal attributes or entities located near leaf regions of the UML graph. For BFS and DFS, traversal was performed over the dual graph representation following \cite{ge2025can} to ensure that all UML associations were included in the final prompt. These transformations modified only the textual presentation order of the UML model and did not alter its underlying semantics or topology.

\begin{table}[t]
\centering
\footnotesize
\caption{Distribution of OCL generation errors across prompting strategies and evaluated LLMs.}
\label{tab:textual_structural}
\begin{tabular}{llrrrrrr}
\hline
Model & Prompting Strategy & Correct & Hal. & Nav. & Log & Inc. & Inv. \\
\hline

\multirow{5}{*}{Claude Sonnet 4.5}
& Classes-first                    & 82 & 2 & 10  & 4  & 8  & 9  \\
& Associations-first               & 80 & - & 12 & 3 & 7 & 13 \\
& Breadth-first search (BFS)       & 81 & 1 & 14 & 3 & 6 & 10 \\
& Depth-first search (DFS)         & 80 & - & 14 & 3 & 4 & 14 \\
& Leaf-first                       & 79 & 1 & 13 & 4 & 5 & 13 \\
\hline

\multirow{5}{*}{GPT-5}
& Classes-first                     & 80 & 5 & 11 & 8  & 6  & 5  \\
& Associations-first                & 80 & 5 & 11 & 2 & 3 & 14 \\
& Breadth-first search (BFS)        & 77 & 1 & 11 & 2 & 6 & 18 \\
& Depth-first search (DFS)          & 77 & 6 & 14 & 2 & 4 & 12 \\
& Leaf-first                        & 82 & 6 & 11 & 4 & 4 & 8 \\
\hline

\multirow{5}{*}{Llama 4 Scout}
& Classes-first                     & 71 & 3 & 20 & 9 & 4 & 8  \\
& Associations-first                & 57 & 6 & 18 & 6 & 4 & 24 \\
& Breadth-first search (BFS)        & 65 & 1 & 17 & 2 & 3 & 27 \\
& Depth-first search (DFS)          & 53 & 3 & 19 & 5 & 7 & 28 \\
& Leaf-first                        & 54 & 3 & 20 & 5 & 7 & 26 \\
\hline

\multirow{5}{*}{Gemini 2.5 Pro}
& Classes-first                     & 79 & 3 & 12 & 8 & 7 & 6 \\
& Associations-first                & 78 & 3 & 10 & 3 & 9 & 12 \\
& Breadth-first search (BFS)        & 73 & 5 & 12 & 2 & 7 & 16 \\
& Depth-first search (DFS)          & 74 & 3 & 12 & 3 & 11 & 12 \\
& Leaf-first                        & 85 & 2 & 11 & 1 & 4 & 12 \\
\hline

\multirow{5}{*}{DeepSeek v3.1}
& Classes-first                     & 30 & 2 & 12 & 3 & 4 & 64 \\
& Associations-first                & 85 & 1 & 12 & 6 & 2 & 9 \\
& Breadth-first search (BFS)        & 81 & 2 & 12 & 5 & 2 & 13 \\
& Depth-first search (DFS)          & 79 & 1 & 7 & 8 & 3 & 17 \\
& Leaf-first                        & 75 & 1 & 18 & 4 & 4 & 13 \\
\hline

\multirow{5}{*}{Grok-4-Fast}
& Classes-first                     & 71 & 8 & 8 & 5 & 10 & 13 \\
& Associations-first                & 82 & 3 & 9 & 4 & 4 & 13 \\
& Breadth-first search (BFS)        & 75 & 6 & 13 & 4 & 3 & 14 \\
& Depth-first search (DFS)          & 78 & 1 & 14 & 5 & 1 & 16 \\
& Leaf-first                        & 72 & 1 & 16 & 4 & 11 & 11 \\
\hline

\end{tabular}
\begin{tablenotes}
\scriptsize
\item Hal: Hallucination, Nav: Navigation, Log: Logic, Inc: Incorrect Ops, Inv: Invalid Ops
\end{tablenotes}
\end{table}

Table~\ref{tab:textual_structural} reports the effect of different textual ordering of the UML class diagram on OCL generation correctness and error distributions across all evaluated LLMs. The results show that LLM performance is sensitive to the textual ordering of UML elements, even though the underlying UML graph remains unchanged. This finding is consistent with prior graph-reasoning studies \cite{ge2025can}, suggesting that LLMs are influenced not only by structural content itself, but also by how that structure is linearized within the prompt.

Across most evaluated models, no single ordering strategy consistently dominated all others. Nevertheless, several important trends emerge. First, traversal-based orderings such as BFS and DFS often increased the number of navigation and invalid-generation errors relative to simpler serialization strategies. For example, GPT-5 produced 11 navigation errors under the classes-first representation, compared to 14 under DFS ordering. Similarly, Claude Sonnet 4.5 exhibited an increase in invalid outputs from 9 under classes-first ordering to 14 under DFS ordering. These results suggest that traversal-based serializations may introduce longer or less locally coherent prompt structures, making it more difficult for the model to maintain consistent navigation paths during OCL generation.

Second, the results indicate that textual order can substantially affect syntactic validity. This effect is particularly visible for DeepSeek v3.1. Under the conventional classes-first representation, DeepSeek achieved only 30 correct generations out of 115 evaluated specifications, while producing 64 invalid OCL expressions. In contrast, under associations-first ordering, correctness increased dramatically to 85 correct generations, with invalid outputs decreasing to only 9. Similar improvements were observed for BFS and DFS orderings. Manual inspection revealed that many invalid generations under the classes-first configuration omitted required OCL context declarations entirely. For example:

\begin{lstlisting}[linewidth=1\linewidth,frame=single]
Project.allInstances()->forAll(p | p.department.locations
->includes(p.location))
\end{lstlisting}

Although structurally plausible, such expressions are not valid OCL invariants because they lack an explicit \verb|context| declaration. One possible explanation is that the classes-first ordering encourages DeepSeek to generate constraints in a more query-like or declarative style rather than following the expected invariant structure. In contrast, traversal-based and associations-first representations may provide stronger relational cues that better align with the model’s learned patterns for structured constraint generation. However, because this behaviour was not consistently observed across the other evaluated LLMs, the exact cause remains unclear and requires further investigation.

The proposed leaf-first ordering produced mixed but noteworthy results. For GPT-5 and Gemini 2.5 Pro, leaf-first ordering achieved the highest correctness among all evaluated strategies, reaching 82 and 85 correct generations respectively. This partially supports the intuition underlying the leaf-first design: presenting structurally terminal classes earlier in the prompt may help the model more easily identify terminal attributes and navigation targets frequently referenced in OCL constraints. However, this behaviour was not universal across models. For Llama 4 Scout and Grok-4-Fast, leaf-first ordering did not improve correctness and in some cases increased navigation-related errors. These results suggest that while leaf-first ordering may help certain models prioritize relevant structural regions of the UML graph, its effectiveness likely depends on how individual LLMs internally process long structured contexts.

More broadly, the results indicate that OCL generation is sensitive to prompt serialization structure even when the semantic content of the UML model remains identical. The observed differences across ordering strategies support the hypothesis that LLMs do not reason over UML graphs in a fully structure-invariant manner. Instead, the textual presentation of structural information itself influences how effectively the model can maintain consistent graph navigation and constraint construction during generation.
\vspace{-0.2cm}
\subsection{RQ3: Prompting Strategies and Error Analysis}

\noindent\textbf{RQ3.1: Effectiveness of Prompting Strategies}

To answer RQ3.1, we evaluated four prompting strategies: zero-shot, Chain-of-Thought (CoT), few-shot, and graph-based prompting. The first three follow standard prompting paradigms for code and formal-language generation, while the graph-based variant explicitly instructs the model to reconstruct the UML diagram as a graph before generating OCL, with the aim of improving structural reasoning and navigation consistency, and reduce graph-related reasoning failures during OCL generation.

For evaluation, a generated OCL constraint was considered \emph{correct} only if it was both syntactically valid and behaviourally consistent with the corresponding gold OCL according to the test-instance-based evaluation procedure described earlier. Incorrect generations were further categorized into five major error types: hallucinated UML elements, navigation errors, logical inconsistencies, incorrect OCL operation usage, and syntactic invalidity. We use the class-first based textual structure when appending the UML class diagram to the prompts.  

\begin{table}[t]
\centering
\footnotesize
\caption{Distribution of OCL generation errors across prompting strategies and evaluated LLMs.}
\label{tab:prompting_errors}
\begin{tabular}{llrrrrrr}
\hline
Model & Prompting Strategy & Correct & Hal. & Nav. & Log & Inc. & Inv. \\
\hline

\multirow{4}{*}{Claude Sonnet 4.5}
& Zero-shot   & 82 & 2 & 10 & 4  & 8  & 9 \\
& Few-shot    & 73 & 4 & 12 & 7  & 6  & 13 \\
& CoT         & 81 & 5 & 9  & 8  & 4  & 8  \\
& Graph-based & 83 & - & 9  & 5  & 6  & 12 \\
\hline

\multirow{4}{*}{GPT-5}
& Zero-shot   & 80 & 5 & 11 & 8  & 6  & 5  \\
& Few-shot    & 82 & 5 & 5  & 8  & 4  & 7  \\
& CoT         & 85 & 5 & 5  & 10 & 4  & 6  \\
& Graph-based & 91 & 1 & 4  & 4  & 5  & 10 \\
\hline

\multirow{4}{*}{Llama 4 Scout}
& Zero-shot   & 71 & 3  & 20 & 9  & 4  & 8  \\
& Few-shot    & 67 & 1  & 18 & 8  & 3  & 16 \\
& CoT         & 69 & 6  & 16 & 11 & 5  & 8  \\
& Graph-based & 60 & 15 & 11 & 5  & 8  & 16 \\
\hline

\multirow{4}{*}{Gemini 2.5 Pro}
& Zero-shot   & 79 & 3 & 12 & 8 & 7 & 6 \\
& Few-shot    & 78 & 7 & 9  & 7 & 7 & 7 \\
& CoT         & 81 & 4 & 13 & 8 & 5 & 4 \\
& Graph-based & 86 & 4 & 15 & 2 & 4 & 4 \\
\hline

\multirow{4}{*}{DeepSeek v3.1}
& Zero-shot   & 30 & 2  & 12 & 3 & 4 & 64 \\
& Few-shot    & 67 & 11 & 10 & 9 & 5 & 13 \\
& CoT         & 41 & 9  & 10 & 8 & 3 & 40 \\
& Graph-based & 80 & 8  & 10 & 3 & 2 & 12 \\
\hline

\multirow{4}{*}{Grok-4-Fast}
& Zero-shot   & 71 & 8 & 8  & 5 & 10 & 13  \\
& Few-shot    & 76 & 5 & 11 & 5 & 8  & 10 \\
& CoT         & 73 & 8 & 9  & 9 & 11 & 5  \\
& Graph-based & 86 & 4 & 12 & 3 & 3  & 7  \\
\hline

\end{tabular}
\begin{tablenotes}
\scriptsize
\item Hal: Hallucination, Nav: Navigation, Log: Logic, Inc: Incorrect Ops, Inv: Invalid Ops
\end{tablenotes}
\end{table}

Table~\ref{tab:prompting_errors} summarizes the correctness and error distributions across prompting strategies and evaluated LLMs. Graph-based prompting consistently achieved the highest or near-highest correctness across most evaluated models. GPT-5 improved from 80 correct constraints under zero-shot prompting to 91 under graph-based prompting, representing the highest overall performance. Similarly, Grok-4-Fast improved from 71 to 86 correct constraints, while DeepSeek-v3.1 exhibited the largest relative improvement, increasing from only 30 correct constraints under zero-shot prompting to 80 correct constraints using graph-based prompting.

The results further show that graph-based prompting substantially reduced several structural reasoning failures. In particular, GPT-5 reduced navigation-related errors from 11 under zero-shot prompting to 4 under graph-based prompting, while hallucination errors decreased from 5 to 1. Similar reductions in navigation and hallucination errors were observed for Claude Sonnet 4.5 and Grok-4-Fast. These findings suggest that explicitly reconstructing the UML model as a graph may help LLMs better maintain structural consistency during OCL generation.

Compared to graph-based prompting, Chain-of-Thought and few-shot prompting produced mixed results. Although CoT prompting occasionally improved correctness, the gains were generally smaller and less consistent across models. Few-shot prompting similarly showed moderate improvements for some models but occasionally introduced additional invalid or hallucinated outputs. For example, Claude Sonnet 4.5 decreased from 82 correct constraints under zero-shot prompting to 73 under few-shot prompting, while invalid outputs increased from 9 to 13.

Despite the overall improvements, graph-based prompting did not uniformly reduce all error categories. Some models continued to exhibit navigation failures and syntactic invalidity even after explicit graph construction. Furthermore, for Llama 4 Scout, graph-based prompting produced lower correctness than the simpler prompting strategies, suggesting that additional structural reasoning steps may increase reasoning instability for certain models.

These results provide partial support for \textbf{H5}. Prompting strategies that explicitly encourage graph construction and structural reasoning generally improve OCL generation performance, particularly for stronger reasoning-oriented models such as GPT-5, Gemini 2.5 Pro, and Grok-4-Fast. However, the effectiveness of graph-based prompting varies substantially across LLM families, indicating that explicit structural decomposition alone is insufficient to fully resolve OCL generation failures.

\noindent\textbf{RQ3.2: Analysis of OCL Generation Errors}

To better understand the limitations of LLM-based OCL generation, we manually analyzed incorrect outputs and categorized the observed failures into five recurring error types: navigation errors, incorrect use of OCL operations, hallucinated model content, logical errors, and invalid or non-OCL constructs. These categories were derived through iterative inspection of generated constraints across all evaluated prompting strategies and models. The results indicate that many failures are closely related to structural reasoning over UML models, rather than purely syntactic generation mistakes. We explain the errors using the class diagram illustrated in Figure \ref{fig:class_diagram_emp}.

\noindent\textit{\textbf{Navigation Errors.}} Navigation errors occur when a generated OCL expression attempts to access model elements through invalid or nonexistent association paths. Although such expressions are often syntactically valid, they violate the structural relationships defined in the UML model and therefore fail during semantic validation or behavioural evaluation. \textit{Example:} Given the natural language requirement, “The start date of an employee as manager of a department must be greater than his/her hire date,” one model generated the following OCL constraint:

\begin{lstlisting}[linewidth=1\linewidth,frame=single]
context Manages inv: self.startDate > self.employee.hireDate
\end{lstlisting}

However, the class \texttt{Manages} contains no association end named \texttt{employee}; the correct navigation should use \texttt{manager}. Consequently, the navigation path \texttt{self.employee} is invalid. This example illustrates how LLMs may incorrectly traverse the UML structure despite producing syntactically plausible OCL. 

\noindent\textit{\textbf{Incorrect Use of OCL Operations.}} This category includes errors in which OCL collection or iterator operations are applied incorrectly. Common examples include invoking collection operators on non-collection values, omitting required iterator expressions, or confusing attribute navigation (\verb|.|) with collection navigation (\verb|->|). \textit{Example:} Given the natural language requirement, “The SSN of employees is an identifier (or key),” the following OCL expression was generated:

\begin{lstlisting}[linewidth=1\linewidth,frame=single]
context Employee inv ssnUnique: Employee.allInstances().SSN->isUnique()
\end{lstlisting}

The \verb|isUnique()| operation requires an iterator expression specifying the uniqueness criterion for each element in the collection. Since no iterator body is provided, the expression is invalid. This example demonstrates that LLMs frequently struggle with the semantic constraints imposed by OCL collection operators.

\noindent\textit{\textbf{Hallucinated Model Content.}} LLMs occasionally generate associations, attributes, or operations that do not exist in the UML model. These hallucinated elements are often semantically plausible but unsupported by the provided diagram, indicating reliance on learned language priors rather than faithful structural reasoning. \textit{Example:} Given the requirement, “The location of a project must be in one of the locations of its department,” the following OCL constraint was produced:

\begin{lstlisting}[linewidth=1\linewidth,frame=single]
context Project inv locationConstraint:
self.controllingDepartment.locations->includes(self.location)
\end{lstlisting}

In this example, the association \verb|controllingDepartment| does not exist in the UML model and is fabricated by the LLM. The generated expression therefore references a nonexistent navigation path.

\noindent\textit{\textbf{Logic Errors.}} Logic errors arise when a generated OCL expression is syntactically valid and structurally consistent with the UML model, but does not correctly encode the intended semantics of the natural language specification. These errors are typically identified during behavioural evaluation using test instances. \textit{Example:} Given the natural language requirement, “The manager of a department must be an employee of the department,” one generated constraint was:

\begin{lstlisting}[linewidth=1\linewidth,frame=single]
context Department inv ManagerIsEmployee: self.manager.department = self
\end{lstlisting}

Although the expression is syntactically valid, it does not correctly enforce the intended constraint. Specifically, the generated OCL only checks whether the manager references the current department, rather than verifying that the manager belongs to the department’s employee collection. Consequently, invalid instances may still satisfy the constraint.

\noindent\textit{\textbf{Invalid or Non-OCL Constructs.}} Some generated expressions contain operations, symbols, or functions that are not part of the OCL language. These errors typically result in parsing failures and indicate that the model is transferring assumptions from general-purpose programming languages into OCL generation. \textit{Example:} Given the requirement, “The location of a project must be in one of the locations of its department,” one model produced the following constraint:

\begin{lstlisting}[linewidth=1\linewidth,frame=single]
context Project inv: self.department.locations.split(',')
->exists(loc | loc.trim() = self.location)
\end{lstlisting}

In this example, the operation \verb|trim()| is not defined in standard OCL. The generated expression therefore contains non-OCL constructs and cannot be executed by the USE validator.

These observed error categories suggest that many OCL generation failures originate from difficulties in maintaining consistent structural reasoning over UML models. In particular, navigation errors and hallucinated associations indicate that LLMs frequently fail to reliably traverse the underlying UML graph, especially when constraints require multi-step reasoning across related classes. While prompting strategies such as graph-based prompting reduce some of these failures, the results show that structural reasoning remains a major challenge for current LLM-based OCL generation systems.
\vspace{-0.2cm}
\section{Discussion}

The results of this empirical study suggest that structural properties of UML class diagrams play an important role in LLM-based OCL generation. In particular, the observed effects of navigation depth, UML complexity, and textual ordering of the UML model indicate that generating correct OCL constraints requires LLMs to maintain consistent reasoning over graph-structured representations of UML models. At the same time, the comparatively weaker and less consistent lexical-similarity effects suggest that OCL generation failures cannot be explained purely through shallow textual matching or token-level associations. These findings motivate a broader discussion on how structural reasoning, graph representation, and prompt organization influence formal specification generation with LLMs.

More broadly, the structural metrics evaluated in this paper represent only a subset of the possible UML properties that may influence OCL generation. While this study focused on navigation depth, class count, association count, and overall model size, many additional graph characteristics may also affect LLM reasoning performance. Examples include node degree distributions, multiplicity constraints, inheritance depth, association types, graph density, centrality measures, and the presence of highly connected hub classes. Similarly, the textual ordering strategies explored in this work represent only a small subset of the possible linearizations of UML graph structures. Since the number of valid serialization strategies grows combinatorially with graph size, exhaustively evaluating all possible orderings is infeasible. Instead, the strategies evaluated here were selected based on prior graph-reasoning literature \cite{ge2025can, han2025reasoning, zhang2024can} and to provide representative examples of structurally different prompt organizations.

Consequently, this work should be viewed as an initial empirical investigation into the relationship between UML graph structure and LLM-based OCL generation. The results establish that structural properties, textual order, and graph-oriented prompting significantly influence OCL correctness, but they also highlight that many aspects of graph-aware reasoning in model-driven engineering tasks remain insufficiently understood. We therefore view this study as a foundation for future work and providing researchers and practitioner direction on graph-sensitive prompting, UML structural analysis, and reasoning-aware evaluation methods for LLM-based formal specification generation.
\vspace{-0.2cm}
\section{Threats to Validity}

\textbf{Construct Validity.}
OCL correctness was assessed using behavioural agreement with gold constraints on manually constructed test instances. While practical, behavioural equivalence on finite cases does not guarantee full semantic equivalence. Correct formulations may be misclassified and vice versa. To mitigate this, test instances included both satisfying and violating cases, including edge scenarios. Additionally, navigation depth was used as a proxy for structural complexity, but it does not capture all aspects of reasoning difficulty (e.g., branching or predicate complexity). Lexical similarity measures were also limited to surface-level string comparisons.

\textbf{Internal Validity.}
Observed relationships between UML properties and correctness do not imply causality, as other factors (e.g., logical complexity) may contribute. Model outputs may also vary due to prompting strategies and stochastic decoding, although consistent settings and repeated sampling were used to reduce this effect.

\textbf{External Validity.}
The study uses the PathOCL dataset (13 models, 115 specifications), which may not represent the full diversity of industrial UML/OCL usage. The focus on class-diagram invariants limits generalizability to more complex or large-scale settings. The small number of models reflects the scarcity of suitable public datasets with aligned UML, natural language, and OCL. Synthetic datasets were avoided due to potential realism and bias concerns, though this limits scale. Future work should validate findings on larger and more diverse datasets.

\textbf{Conclusion Validity.}
The use of GEE accounts for correlated observations, but the small number of UML models may limit statistical power. Results were consistent across sensitivity analyses.

\section{Related Work}

Automatically generating OCL constraints from natural language specifications has been studied widely within model-driven engineering research \cite{cabot2022combining}. Early approaches primarily relied on rule-based transformations, template matching, and syntactic parsing techniques to translate restricted forms of natural language into OCL expressions \cite{bajwa2010ocl, tan2010ocl}. These approaches typically required carefully structured input specifications and were limited in their ability to generalize across domains and linguistic variations.

More recent work has explored the use of Large Language Models (LLMs) for OCL generation \cite{siala2025using, li2025optimizing, yang2024deepocl}. Abukhalaf et al.~\cite{abukhalaf2023codex} investigated prompt engineering strategies for Codex-based OCL generation and demonstrated that prompting design significantly influences generation quality. PathOCL \cite{abukhalaf2024pathocl} further introduced path-based prompt augmentation strategies intended to improve structural navigation during OCL generation. Other recent work has explored dataset construction and fine-tuning approaches for improving LLM-based OCL generation \cite{pan2024generative, mayr2025generating, li2025optimizing}. These studies primarily focus on improving generation accuracy through prompting or training strategies.

Furthermore, recent research has shown that LLMs struggle with graph reasoning tasks when graph structure is represented in natural language form \cite{tsitsulin2024graph, zhangimproving, zhang2024can, han2025reasoning, yuan2025ma}. The NLGraph benchmark \cite{wang2023can} demonstrated that LLM performance degrades substantially on tasks such as shortest-path reasoning, connectivity analysis, and Hamiltonian path problems as graph complexity increases. Subsequent work further showed that LLM reasoning is highly sensitive to graph serialization and descriptive ordering \cite{ge2025can}.

Other studies have proposed methods for improving graph reasoning capabilities \cite{chai2025graphllm}. GraphOtter \cite{li2025graphotter} explored graph-evolving reasoning strategies for complex reasoning tasks, while Peng et al.~\cite{peng2025rewarding} investigated reinforcement-learning-based approaches for improving graph reasoning generalization. These studies collectively suggest that current LLMs often struggle to maintain consistent structural reasoning over multi-hop graph relationships.

Additionally, representing UML class diagrams as graphs has been widely studied in model-driven engineering and formal verification research. Prior work has explored transformations from UML models into graph transformation systems for analysis and verification purposes \cite{holscher2006translating, ziemann2005uml, alsammak2026effective}. These graph-based representations enable reasoning about structural relationships, navigability, and model consistency. 

OCL validation itself has also been extensively studied \cite{soeken2010verifying}. USE \cite{gogolla2007use} provides executable validation support for UML and OCL models, while other work has explored constraint solving and automated test-data generation for OCL specifications \cite{ali2013generating, franconi2019ocl, cabot2006constraint}.

In contrast to prior work, we investigates why LLMs fail in OCL generation by linking these failures to limitations in structural reasoning over UML graphs. We show that common errors, such as invalid paths and hallucinated associations, mirror known graph reasoning issues, but within a formal software engineering context, highlighting how UML structural properties impact LLM performance.

\section{Conclusion and Future Work}

This paper examined why Large Language Models (LLMs) struggle to generate correct OCL constraints from natural language and UML class diagrams. We approached this problem from a graph-reasoning perspective, analyzing how UML structure influences generation correctness.

Using the PathOCL dataset, we evaluated multiple LLMs and analyzed the impact of navigation depth, structural complexity, lexical similarity, and prompting strategies. Results show that correctness declines as navigation depth and UML complexity increase, with multi-hop traversals posing significant challenges. These findings align with known limitations of LLMs in graph reasoning.

Lexical factors played a limited role, indicating that failures stem more from structural reasoning difficulties than superficial text matching. While graph-aware prompting improved performance slightly, common errors persisted, including invalid navigation, hallucinated associations, and logical inconsistencies. OCL generation is not merely a translation task but a graph reasoning problem. Improving performance will likely require advances in structural reasoning, such as graph-aware and neuro-symbolic approaches, rather than prompt engineering alone.

\section{Data Availability}
To support reproducibility and comply with the conference open science policy, the datasets, prompts, and evaluation artifacts used in this study have been made publicly available at DOI: https://doi.org/10.5281/zenodo.20279636.

\bibliography{references}

\end{document}